# Photonic Maxwell's demon as an energy resource

M. G. Raizen, A. Barr, and A. Helal *
Dept. of Physics, University of Texas at Austin, Austin, TX 78712, USA

We propose a new realization of Maxwell's demon for photons and its use as an energy resource. The starting point is a multilayer dielectric mirror that transmits photons within a certain wavelength band but reflects photons at longer wavelengths, followed by a material that irreversibly converts the wavelength. A 3D box constructed from these one-way mirrors is a self-acting Maxwell's demon for photons. The first case considered is a one-way mirror for visible photons, converting blue to red as an example. The next case is a one-way mirror that transmits visible light but reflects infrared light, followed by a perfect blackbody. We analyze a 3D box with such mirrors as walls to accumulate and store blackbody radiation. Applications include solar-powered thermal energy and solar-powered thermophotovoltaics. The latter case would combine the 3D box with an ultra-low bandgap semiconductor photovoltaic cell inside. In the ideal optical limit, radiation can enter the box but cannot return through the entrance channel after conversion, allowing repeated photon recycling and nearly complete optical utilization. This can in principle move the conversion efficiency toward the thermodynamic limit for solar-energy conversion, the Landsberg limit of 93.3%.

## I. INTRODUCTION.

In 1871, James Clerk Maxwell proposed a thought experiment which is known as Maxwell's demon [1–3]. He envisioned a box containing molecules in gas phase with a partition separating it into two chambers. An intelligent creature, the demon, could observe the motion of each molecule, and open a trap door to let a molecule through while keeping the others out. This construction could move all the molecules in the box to one of the chambers without doing any work, in apparent violation of the second law of thermodynamics. The conceptual resolution to the paradox was that the demon collected information, carrying away entropy and avoiding any violation of the second law [4]. In recent years, several experimental realizations of Maxwell's demon have been reported. These include atomic ensembles, described in more detail below [5–9], single electrons [10], molecular ratchets in solution [11], and colloidal particles [12–15].

The common interpretation of Maxwell's demon implies measurement and feedback control of individual particles, such as single electrons and colloidal particles. Feedback control of a single particle in a box was first proposed by Leo Szilard as a way to make a heat engine based upon information [16]. This textbook concept and subsequent experimental realizations elucidate the physics of Maxwell's demon, but do not appear to have any real-life applications. The real challenge is to control large ensembles, a task that Maxwell deemed "impossible to us." His proposed solution was a self-acting device, which would avoid the need for active measurement and feedback.

In 2005, we proposed a one-way wall for atoms in gas phase based on a conservative potential combined with the irreversible process of laser optical pumping which induces spontaneous decay of atoms as they cross the wall [5]. This construction breaks reciprocity in a simple and intuitive manner and is an exact realization of Maxwell's idea of a self-acting device. Although the entropy of the atoms decreases, there is no violation of the second law of thermodynamics due to the increase in information entropy as atoms cross the one-way wall. The experimental validation in our group followed a few years later [6–8], while another experiment used an all-optical potential instead of a magnetic potential [9]. The one-way wall is the resolution to Maxwell's challenge for large ensembles, so the problem is solved for atoms. This is not just a pedantic point; controlling large ensembles of atoms is proving to be an efficient method for separating isotopes of the elements and will have an important impact on medicine [17,18].

Building on this work, we propose extending Maxwell's demon to photons, inspired by the prior realization with atoms. A photonic Maxwell's demon has also been experimentally realized using measurement at the few-photon level followed by feed-forward to extract work from thermal light [19]. This information-based implementation is conceptually distinct from the self-acting frequency-converting wall considered here, which requires neither measurement nor feedback.

The outline of this paper is as follows. We first propose a one-way mirror for photons as a self-acting device. We then use the new construct to form a 3D photon-trapping box and show how, in the ideal limit of vanishing return loss, it can approach the thermodynamic maximum efficiency allowed by the incident radiation. Our first application is to convert sunlight into thermal energy for heating of a working fluid. Our second application is to convert sunlight into electrical power using a low-bandgap photovoltaic cell inside the photon-recycling box. Detailed thermodynamic benchmarks and photon-recycling calculations are retained in Appendices A-C.

* Contact author: ahelal@utexas.edu

## II. SELF-ACTING ONE-WAY MIRROR FOR PHOTONS

We start with optics and consider how to make a one-way mirror. Most optical elements are reciprocal in nature, including mirrors, beam splitters and lenses. An exception is the Faraday isolator, in which magneto-optic rotation does not reverse upon reversal of the propagation direction; together with polarizers this produces nonreciprocal transmission. It is interesting to ask whether one could develop more general methods to break reciprocity, which can violate the conventional directional form of Kirchhoff's law of radiation. Indeed, this question has been explored using metamaterials, a major area of modern research. Non-reciprocal photonic structures have been proposed and demonstrated, motivated by the interest in passive cooling with flat panels. These systems are mainly based on magnetic materials, Weyl/topological materials, and active spatiotemporal modulation [20].

However, if we want to use a one-way mirror to construct Maxwell's demon, nonreciprocity alone is not sufficient; an irreversible change of state is also required. The approach considered here is therefore different. The dielectric mirror itself may remain reciprocal, while absorption followed by spontaneous or thermal re-emission changes the photon frequency and makes the process irreversible. The forward and reverse paths therefore involve different photon states. This is analogous to optical pumping followed by spontaneous emission in the atomic one-way wall [21].

To gain intuition, we return to the one-way wall for atoms [5] where they absorb photons from a laser in their ground electronic state and spontaneously decay to a different internal state. In this process, the atoms convert the wavelength of the incident laser, typically to a longer wavelength, as they decay. This mechanism provides a way to construct a one-way mirror for visible light. Consider a dielectric mirror that transmits blue light but reflects red light, a standard optical element known as a dichroic mirror.

While commercial dichroic mirrors are angle dependent, omnidirectional dichroic mirrors can be designed and fabricated using multilayer dielectric materials. The key development over the past few years has been an efficient computational approach with machine learning; it starts with the desired reflectivity as a function of wavelength, and then computes the required combination of dielectrics, which typically comprises hundreds of layers. The mirrors produced in this way can be omnidirectional, polarization insensitive, and with reflectivity higher than 99% in a desired band [22,23].

Now imagine that one side of the mirror is coated with a material that absorbs in the blue and emits in the red, such as europium-doped phosphors that are used in light emitting diodes (LEDs). This configuration is illustrated in FIG. 1(a). Blue light from the left goes through the dichroic mirror and then encounters the phosphor where it is absorbed and converts to red light. This light cannot return to the left due to the reflectivity of the dichroic mirror and therefore can only propagate to the right. Blue light from the right first encounters the phosphor where it is converted to red light that cannot go through the dichroic mirror. This combined system is a one-way mirror for blue light and works by irreversible conversion of blue to red. A 3-D box to store light is illustrated in FIG. 1(b). Maxwell's demon for visible light has interesting applications to nonlinear optics which will be discussed in a later publication.

The idealized discussion assumes high absorption of the incident blue light, high phosphor quantum yield, negligible red self-absorption, and high red-band reflectivity. Finite conversion efficiency, parasitic absorption, or red leakage makes the one-way action probabilistic rather than perfect.

## III. BLACKBODY PHOTON TRAPPING AND SOLAR THERMAL ENERGY

Now, we apply the same ideas to convert visible light to the infrared. A multi-layer dielectric mirror can be constructed that is highly reflective over the blackbody spectrum, except the visible part of the spectrum where it is highly transmissive. To make this into a one-way mirror, one just adds a nearly perfect blackbody that absorbs the visible light which is re-emitted over the rest of the blackbody spectrum [24]. We claim that this configuration realizes Maxwell's demon near maximum efficiency, which has never been proposed or realized to the best of our knowledge, primarily limited by the dielectric mirrors [25].

The operation of the device is simulated numerically, and the results are shown in FIG. 2. The temperature is shown as a function of time for different values of mirror reflectivity across the blackbody spectrum. Increasing the infrared reflectivity substantially raises the equilibrium blackbody temperature because a progressively smaller fraction of the internally generated thermal radiation escapes from the cavity. The simple wavelength-selective implementation should not, however, be identified with an ideal one-way thermodynamic boundary at arbitrarily high temperature: the blackbody develops a short-wavelength tail in the same spectral band that must remain transmissive to admit sunlight.

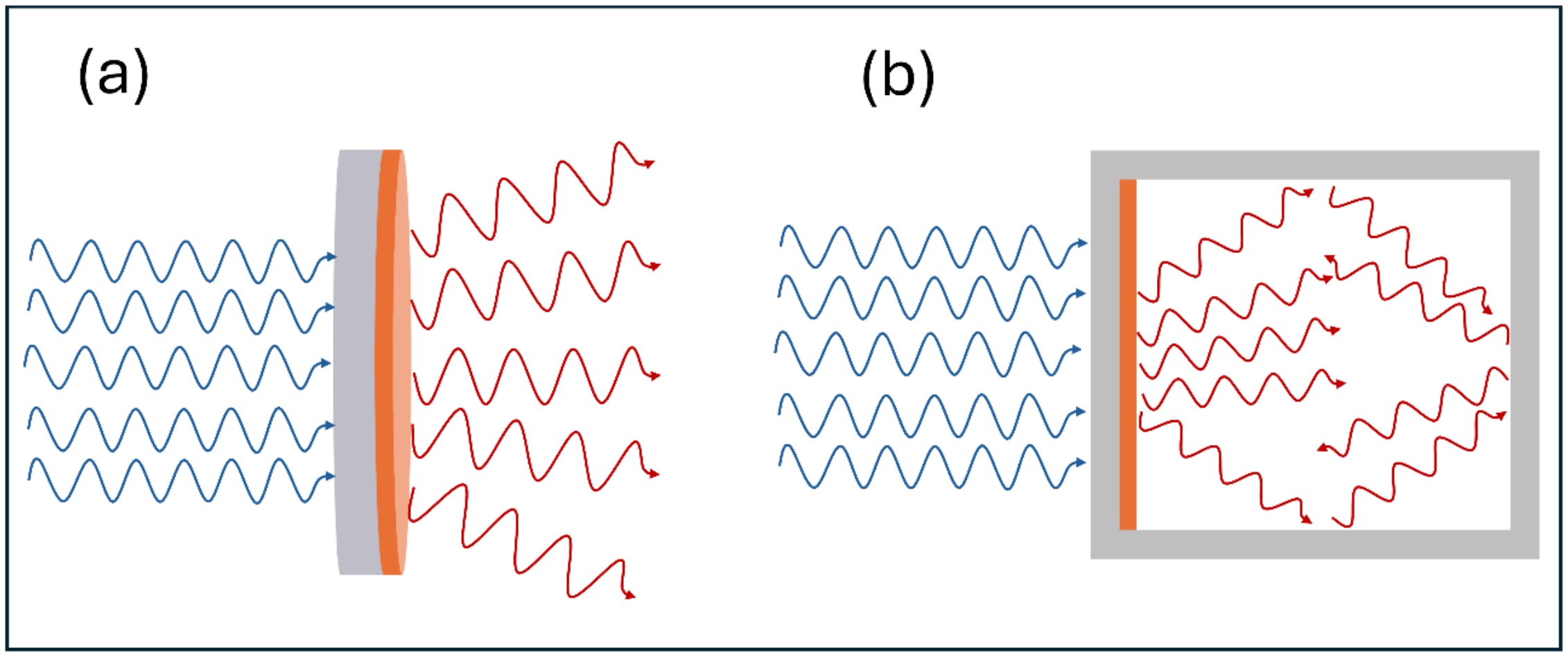

FIG. 1. (a) A one-way mirror consists of a dielectric mirror that transmits blue light and reflects red light (grey) and a phosphor coating that converts blue light to red (orange). (b) A 3D box is constructed (illustrated in 2D) that accumulates and stores red light.

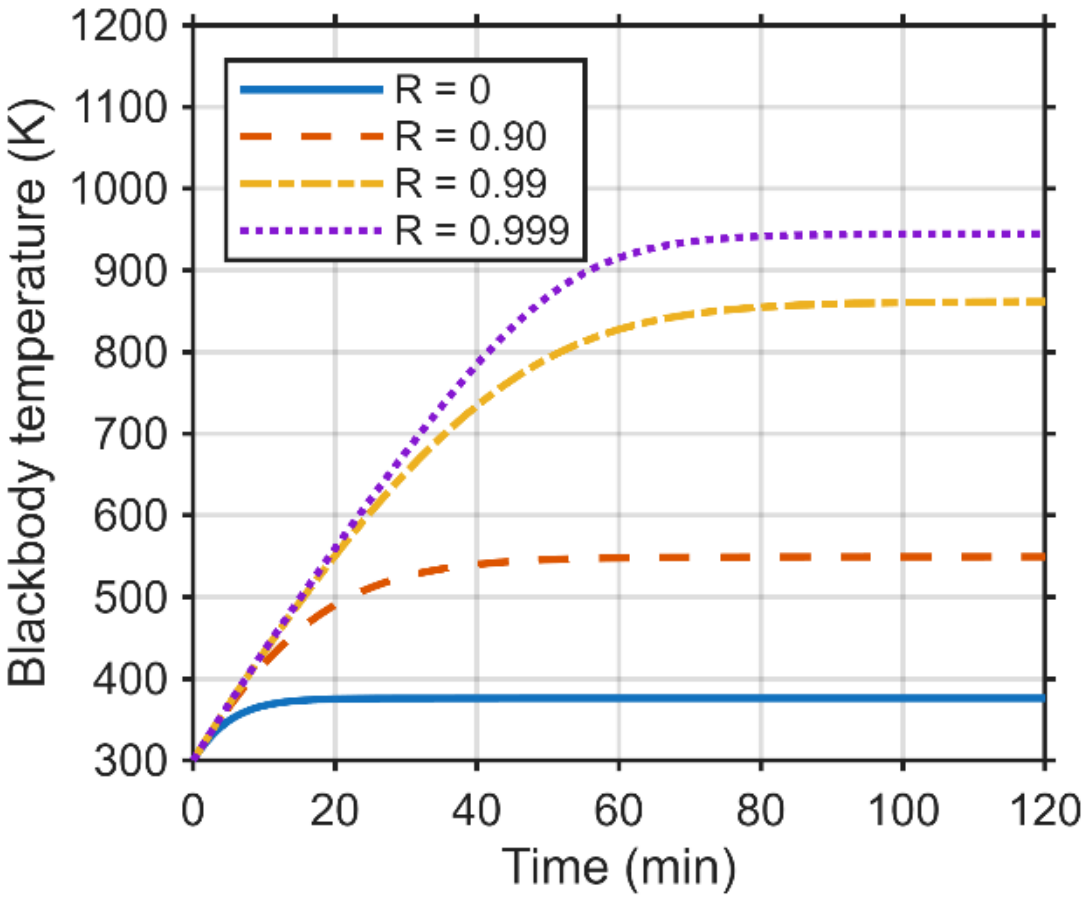

FIG. 2. Blackbody temperature as a function of time for different values of mirror reflectivity. The time is determined by the heat capacity, which in this case is assumed to be a carbon-coated stainless-steel sheet of thickness 1 mm. A thickness of 10 micrometers leads to heating on the time scale of seconds.

We now assume that a liquid (e.g. water) flows in and out of another blackbody inside the box. At the input, the water is at room temperature, and at the output it is heated by the blackbody inside the box. The flow rate can be adjusted to obtain a desired output water temperature, such as 550 K which is used for geothermal energy. In the ideal limit, in which nearly all of the incident sunlight is admitted and retained in the box, the available thermal power can approach the incident solar irradiance, approximately 1000 W/m$^2$ at peak illumination.

The most likely application would be for single-home heating, though large scale-ups should be possible due to the low cost of the method as compared with drilling deep wells for geothermal energy. In this application the useful output is thermal power rather than electrical work, and the governing requirement is that absorbed solar power exceeds radiative and parasitic losses at the desired operating temperature. The box therefore provides a direct experimental route for testing how much the frequency-converting wall suppresses radiative return in practice. Related solar-thermal concepts using an infrared mirror to suppress reradiation have also been analyzed [25]; the distinction here is the proposed frequency-converting one-way boundary.

## IV. THERMOPHOTOVOLTAIC ENERGY CONVERSION

Another application is for thermophotovoltaics, devices that convert heat directly into electricity in a low bandgap semiconductor. Thermophotovoltaics are currently designed to convert waste heat from power plants or other industrial settings and have reached very high efficiency [26,27]. We propose that a photonic Maxwell's demon applied to thermophotovoltaics could efficiently convert solar energy into electricity. The configuration would be similar to the above case of solar thermal energy, but instead of a flowing liquid to remove heat, infrared photons would generate electricity in a semiconductor photovoltaic cell.

The semiconductor in this case should have an ultra-low bandgap, so that the transmission window of visible light could be extended. For example, InN was found to have a bandgap of 0.7 eV, though p-doping has been a challenge for incorporation into photovoltaics [28]. More recently, another material, InAs/InAsSb, was shown to have a bandgap below 0.25 eV [29]. The proposed application of low and ultra-low bandgap semiconductors to solar energy was to stack multiple photovoltaics in tandem to cover the entire solar spectrum. This approach was very difficult to implement in practice due to many factors, such as lattice mismatch. Here, we present a different and counter-intuitive solution: use a single photovoltaic with an ultra-low bandgap inside the photon-recycling box, where repeated recycling can strongly increase the fraction of the trapped radiation that is eventually converted to electricity [30].

To provide an example of our new approach, the fraction of sunlight with a wavelength less than 1600 nm is about 90%. This incoming visible light would be converted to blackbody radiation inside the box, and even at a temperature of 1200 K, only 5.5% of the blackbody radiation lies below 1600 nm and could escape out through the mirrors. Photons within a certain band would be absorbed by the semiconductor and converted into electricity, while photons outside the band could be reflected back to the blackbody by another specially designed dielectric mirror to avoid parasitic absorption inside the photovoltaic cell. These reflected photons are not necessarily lost; they can be reabsorbed by the blackbody, re-emitted, and given additional opportunities to interact with the photovoltaic cell. In the ideal Maxwell-demon limit, where the converted radiation cannot return through the entrance channel, repeated photon recycling can drive the optical utilization toward unity, while the ultimate solar-to-work efficiency remains bounded by the solar exergy limit.

Again, we turn to atomic physics for intuition: the method of evaporative cooling of atoms which is used to produce Bose-Einstein condensates [31]. In this method, an ensemble of atoms is typically confined to a magnetic trap, where they undergo elastic collisions. The hottest atoms are then ejected from the trap by flipping their spin, leaving a non-equilibrium velocity distribution. The colliding atoms reestablish thermal equilibrium at a lower temperature, filling in the truncated velocity tail. In the case of atoms, the particle number is conserved, so evaporative cooling entails a loss of atoms as they cool. If we now consider the efficiency of thermophotovoltaics in a Maxwell's demon box, absorption of photons within the semiconductor band will create a non-equilibrium distribution of photons in the box, but absorption and emission will reestablish a blackbody spectrum. Unlike atoms, the photon number in the box is not conserved, but the crucial point is that the blackbody itself is the photon converter.

We have simulated the thermalization process, and the resulting equilibrium blackbody temperature is shown in Fig. 3 as a function of the TPV bandgap and the infrared reflectivity of the solar-entrance dichroic. The calculation shows that suppressing radiative return through increased reflectivity substantially raises the equilibrium cavity temperature, while the TPV bandgap controls the spectral channel through which energy is extracted. Re-thermalization of the photon gas is irreversible and therefore generates entropy. This is not a flaw in the Maxwell-demon construction; it is the mechanism that changes the photon state. For energy harvesting, however, that entropy generation must be included in the exergy balance.

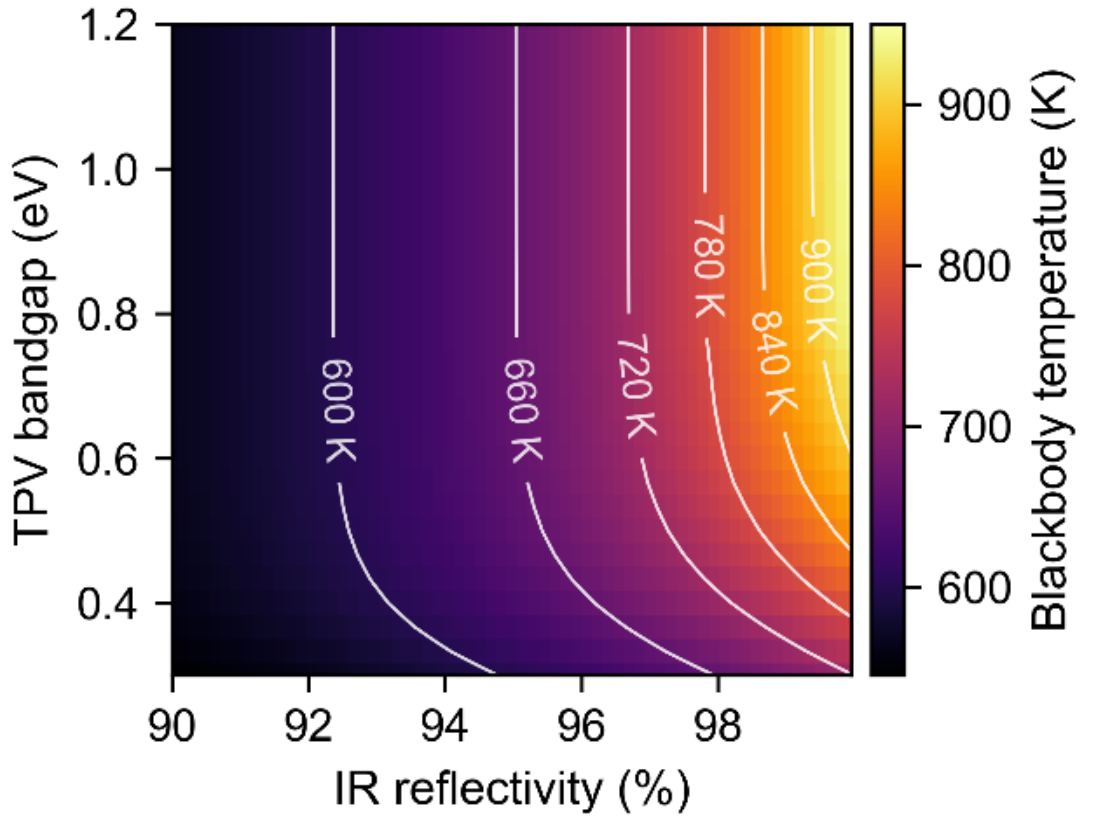

FIG. 3. Equilibrium blackbody temperature as a function of TPV bandgap in eV and IR reflectivity of the solar entrance dichroic. The TPV is contained in a cavity attached to the blackbody cavity from FIG. 2, separated by a second dichroic which admits an infrared wavelength band equivalent to 0.05 eV centered around each TPV bandgap value.

The relevant thermodynamic hierarchy is known from positive-illumination solar-energy conversion [32–34]. Exergy is the maximum useful work that can, in principle, be extracted from a system as it comes into equilibrium with its environment [35]. For the present passive architecture, the maximum extractable work is therefore determined by the exergy balance associated with irreversible frequency conversion:

$$\dot{W} \le \dot{X}_{Solar} - \dot{X}_{leak} - T_a \dot{S}_{gen}, \quad \dot{S}_{gen} \ge 0 \qquad (1)$$

where $\dot{X}_{solar}$ is the exergy admitted from the solar radiation, $\dot{X}_{leak}$ is the exergy that escapes through unwanted optical or parasitic channels, and $\dot{S}_{gen}$ is the entropy generated by irreversible spectral conversion, absorption, re-emission, and other dissipative processes. Equation (1) states the physically testable requirement for an energy advantage: suppressing radiative return must save more exergy than is destroyed by the added irreversible conversion process.

For a 6000 K source and a 300 K environment, the conventional one-stage blackbody limit is 85.36%, the reciprocal multicolor limit is 86.8%, and the ultimate solar-radiation exergy limit is the Landsberg value of 93.33% [32]. An ideal two-stage radiative cascade gives about 90.03% and the optimized multistage limit approaches 93.33% as the number of stages increases. These are thermodynamic benchmarks, not efficiencies assigned to the passive frequency-converting wall. The detailed calculation is given in Appendix A.

The optical role of the Maxwell-demon boundary can be separated from this thermodynamic limit. Finite wall reflectivity controls how long trapped photons remain available for useful conversion. For a useful-converter encounter probability $g = 0.01$, the recycling model in Appendix C gives a useful-capture probability of approximately 91% at $R = 0.999$ and approximately 99% at $R = 0.9999$. If all other losses are deliberately excluded, these correspond to optical-loss-only work ceilings of about 84.9% and 92.4%, respectively; as R approaches unity, the wall-loss probability approaches zero and this deliberately optimistic ceiling approaches the 93.3% Landsberg benchmark. Appendixes B and C retain the full photon-survival and recycling calculations.

## V. DISCUSSION AND CONCLUSION

The ideal Maxwell-demon statement can therefore be made precisely: the boundary may be lossless with respect to optical return, so that energy admitted into the trapped spectral state cannot escape through the entrance channel. This can drive optical utilization toward unity through repeated recycling. It does not follow that the same device converts 100% of the incident solar energy into work. Even when optical escape is eliminated, the incoming radiation carries entropy, and the irreversible state-changing process used to realize the passive wall may generate additional entropy. The theoretical solar-to-work maximum is consequently the Landsberg exergy limit in the ideal zero-additional-entropy benchmark and is lower when $\dot{S}_{gen} > 0$.

For thermophotovoltaics, the practical converter can be optimized separately once the wall has been characterized. The calculations in Appendixes B and C quantify only the optical advantage supplied by trapping and repeated encounters; they do not constitute a detailed photovoltaic-device model. A physical passive frequency-converting wall will also be affected by finite spectral leakage, mirror absorption, imperfect conversion efficiency, angular and polarization dependence, and entropy generation during absorption and re-emission. These effects can only lower the attainable work from the ideal benchmark.

To conclude, we propose a self-acting Maxwell's demon for photons based on spectral selectivity and irreversible frequency conversion. In the ideal optical limit, the wall admits radiation into the cavity but prevents the converted radiation from returning through the entrance channel. Repeated photon recycling then drives the useful-interaction probability toward unity and can eliminate the corresponding optical escape penalty. This is the central advantage of the Maxwell-demon boundary. The result does not imply 100% solar-to-work conversion: established thermodynamic analysis places the ultimate solar-radiation work limit at 93.3% for a 6000 K source and 300 K environment [32–34]. Our calculations show how finite reflectivity moves the optical-loss-limited ceiling below that value and how $R \rightarrow 1$ restores the Landsberg benchmark. Demonstrating experimentally that the one-way boundary suppresses optical return while keeping the additional penalties small would establish the energy-conversion value of the photonic Maxwell demon. Thus, Maxwell's demon can be realized for photons and may provide a new route toward high-efficiency energy conversion. Further applications are likely to emerge once the proposed methods are validated experimentally. More generally, Maxwell's demon for photons may enable a new level of control over electromagnetic radiation.

## ACKNOWLEDGMENTS

This work was supported by the Sid W. Richardson Foundation.

## APPENDIX A: THERMODYNAMIC EFFICIENCY BENCHMARKS

The thermodynamic limits used in the main text can be obtained from the positive-illumination solar-energy analysis of Ref. [32]. For an intermediate blackbody at temperature T, with ideal Carnot work extraction to an environment at temperature $T_a$, the one-stage efficiency is

$$\eta_1(T) = \left[1 - \left(\frac{T}{T_s}\right)^4\right]\left(1 - \frac{T_a}{T}\right) \quad \text{(A1)}$$

For $T_s$ = 6000 K and $T_a$ = 300 K, Eq. (A1) is maximized at T ≈ 2544 K, giving $\eta_1$ = 85.36%, the conventional blackbody limit. Buddhiraju et al. also obtain the reciprocal multicolor limit of 86.8% when the incident spectrum is separated into independently optimized spectral channels [32]. These limits are properties of particular thermodynamic architectures, not of the photovoltaic material alone.

The ultimate bound is set by the exergy of the incident solar radiation. For blackbody radiation the Landsberg efficiency is

$$\eta_L = 1 - \frac{4}{3}\left(\frac{T_a}{T_s}\right) + \frac{1}{3}\left(\frac{T_a}{T_s}\right)^4 = 93.33\% \quad \text{(A2)}$$

The difference between the reciprocal blackbody limit and the Landsberg limit is associated with irreversibility and radiative return. In the ideal multistage construction of Ref. [32], broadband circulators redirect the emission of each intermediate blackbody to a lower-temperature stage instead of allowing it to return directly toward the Sun. For N-stages the corresponding ideal benchmark is

$$\eta_N = \sum_{i=1}^{N}\left[\frac{\left(T_{i-1}^4 - T_i^4\right)}{T_s^4}\right]\left(1 - \frac{T_a}{T_i}\right), \quad T_0 = T_s \quad \text{(A3)}$$

Optimizing the intermediate temperatures gives $\eta_1$ = 85.36%, $\eta_2$ ≈ 90.03%, and a monotonic approach to the Landsberg limit as N increases. Ghalekohneh and Zhao showed that an ideal nonreciprocal solar-thermophotovoltaic intermediate layer with unit front-side absorptivity, vanishing front-side emissivity, and emission routed to the converter can likewise reach the Landsberg limit [33]. Related nonreciprocal solar and photovoltaic architectures reach the same class of upper bounds [34].

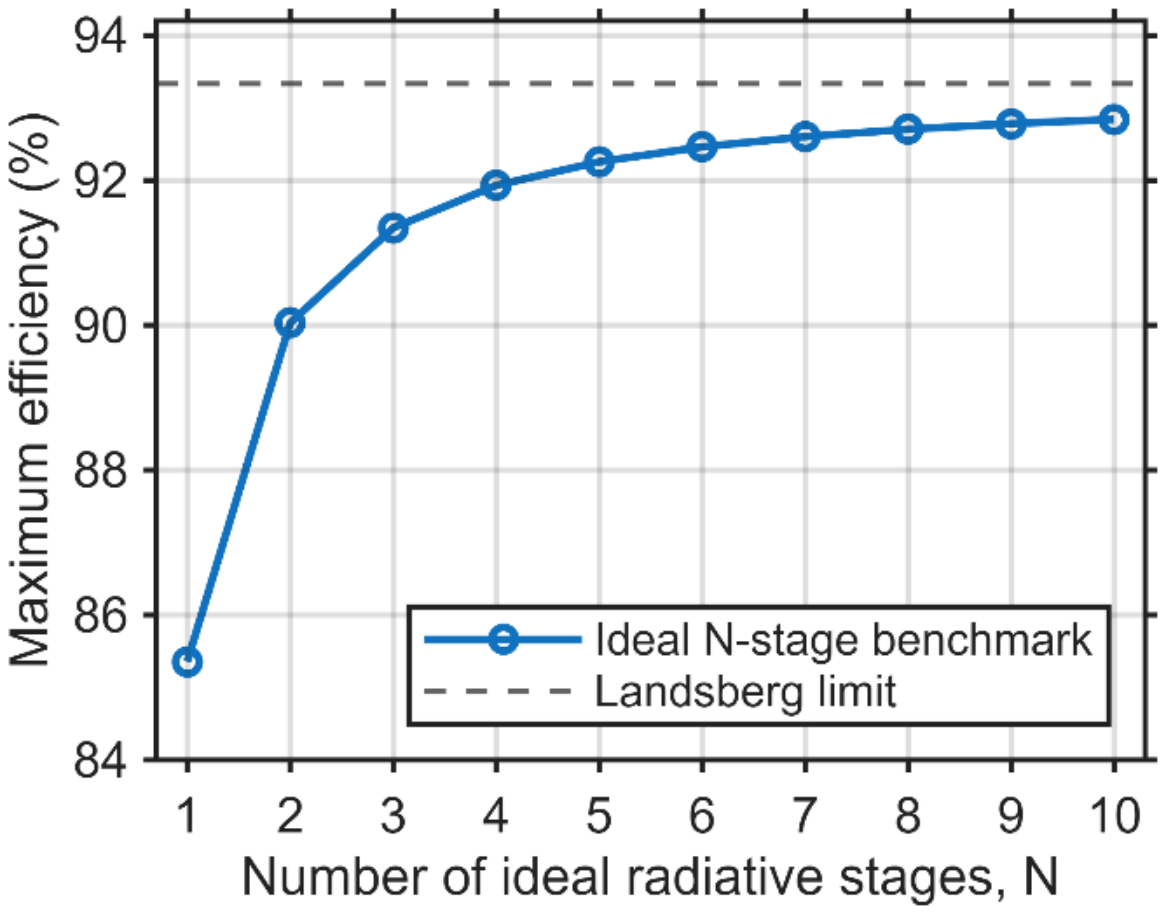

FIG. 4. Thermodynamic benchmark for ideal one-way radiative routing. The optimized N-stage blackbody/Carnot cascade gives 85.36% for N=1, 90.03% for N=2, and approaches the 93.33% Landsberg limit as N → ∞ [32].

Figure 4 is therefore a benchmark for ideal unidirectional radiative routing. It is not a prediction that the passive frequency-converting wall proposed here is equivalent to an ideal circulator. For the passive architecture, the appropriate statement remains the exergy balance in Eq. (2): the benefit of suppressing radiative return must exceed the exergy destroyed by the irreversible conversion process.

## APPENDIX B: FINITE-REFLECTIVITY PHOTON TRAPPING

To isolate the effect of photon trapping from the separate thermodynamic conversion problem, consider a trapped-band photon that encounters a wall of reflectivity R. If parasitic wall absorption or escape is the only optical loss at each encounter, the fraction of the original photon population remaining after N wall encounters is

$$P_{surv}(N) = R^N, \quad \text{(B1)}$$

Thus, increasing R does not create energy, but it increases the residence time and the number of opportunities for the photon to encounter a useful absorber or converter. In the ideal Maxwell-demon return boundary, R = 1 for the converted or trapped state and $P_{surv}$ = 1 for any number of encounters: the return-channel optical loss is zero.

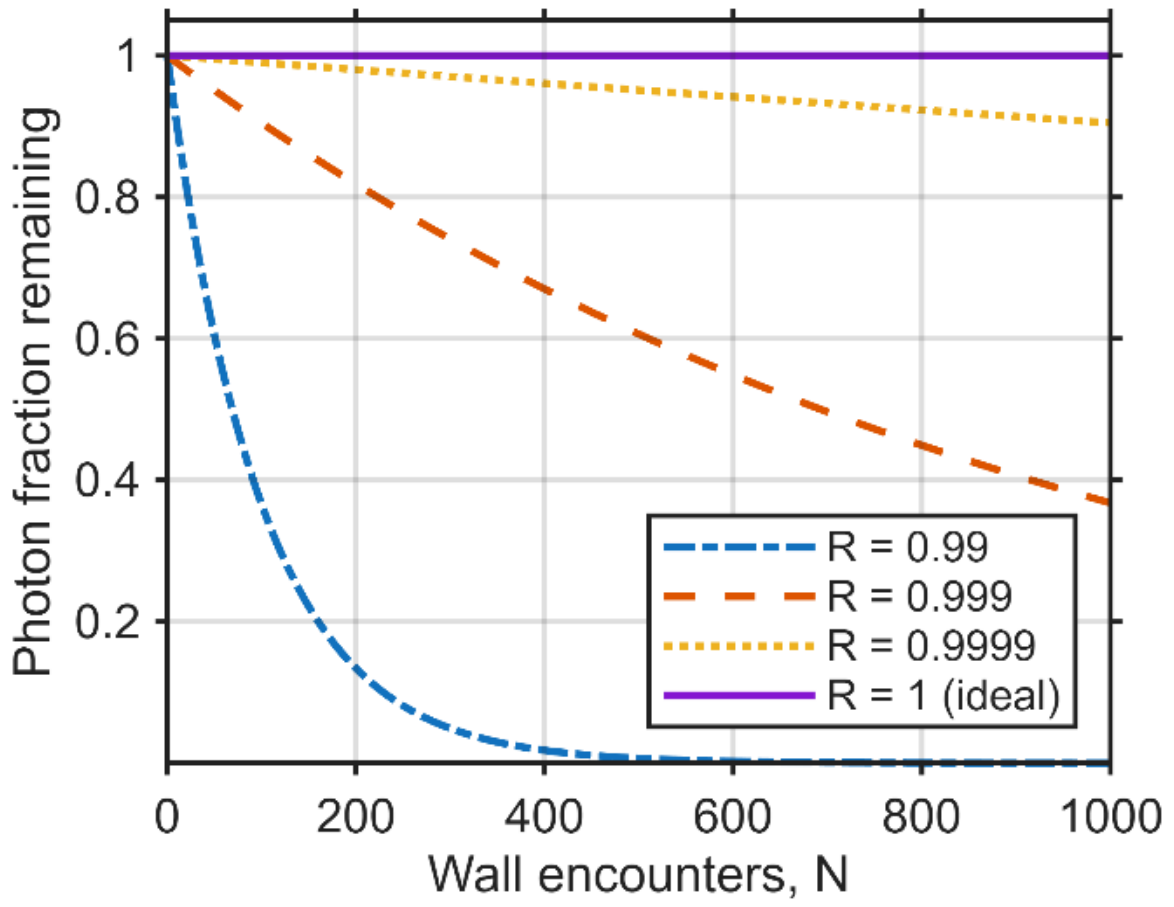

FIG. 5. Photon trapping versus wall reflectivity. The surviving photon fraction after N wall encounters is $R^N$; as R →1, photon loss is progressively suppressed, with the ideal Maxwell-demon limit R=1 retaining the trapped population.

Figure 5 illustrates how small per-encounter losses accumulate over many reflections. For R = 0.99, the trapped photon population decays rapidly, whereas progressively higher reflectivity greatly extends the photon residence time. At R = 0.9999, approximately 90% of the population remains after 1000 wall encounters, while in the ideal limit no photons are lost through this channel. Thus, very high reflectivity becomes increasingly important when useful conversion relies on many recycling events.

## APPENDIX C: REPEATED PHOTON RECYCLING AND OPTICAL UTILIZATION

A second simple model quantifies the benefit of repeated recycling when a useful converter occupies a fraction g of the effective cavity encounter probability. At each surface encounter the photon reaches the converter with probability g; otherwise, it reaches the wall, where it is reflected with probability R or lost with probability 1 − R. Summing the geometric series gives

$$P_{use} = \frac{g}{1-(1-g)R}, \qquad P_{loss} = \frac{(1-g)(1-R)}{1-(1-g)R}, \tag{C1}$$

The useful-conversion and loss probabilities satisfy $P_{use} + P_{loss} = 1$. As the mirror reflectivity approaches unity, repeated photon recycling increases the probability that an admitted photon eventually interacts with the converter. For example, with g = 0.01, $P_{use}$ increases from approximately 0.910 at R=0.999 to 0.99 at R=0.9999. In the ideal lossless limit, $R = 1$, any nonzero encounter probability gives $P_{use}$=1, because photons that are not converted on one encounter remain available for subsequent encounters.

An important consequence of Eq. (C1) is that the photovoltaic area need not be comparable to the solar-collection area. In a well-mixed cavity, reducing the photovoltaic area lowers the probability of interaction on any single pass, but this reduction can be compensated by repeated recycling. For finite reflectivity, the required cell area is therefore governed by the competition between the encounter probability and the number of recycling events available before optical loss. In practice, the photovoltaic area cannot be reduced indefinitely, because directing the same radiative power onto a smaller cell increases the demands on thermal management and electrical transport. Nevertheless, this geometry could allow an expensive ultra-low-bandgap photovoltaic to occupy substantially less area than the solar collector.

To connect this purely optical result to the thermodynamic discussion without introducing a detailed TPV device model, we define an optimistic optical-loss-only work ceiling

$$\eta_{max,\,opt}(R,g) = \eta_L P_{use}, \tag{C2}$$

Equation (C2) assumes that finite wall escape is the only additional loss and that all radiation reaching the useful converter can otherwise be processed up to the solar-exergy limit. It is therefore an upper bound, not a prediction for the real phosphor or blackbody wall. For g = 0.01, the ceiling rises from about 84.9% at R = 0.999 to 92.4% at R = 0.9999 and approaches 93.3% as R approaches unity. The corresponding parasitic wall-loss probability falls from about 9.0% to about 1.0%. This calculation makes precise the sense in which photon trapping can eliminate an avoidable optical loss channel while leaving the Landsberg bound intact.

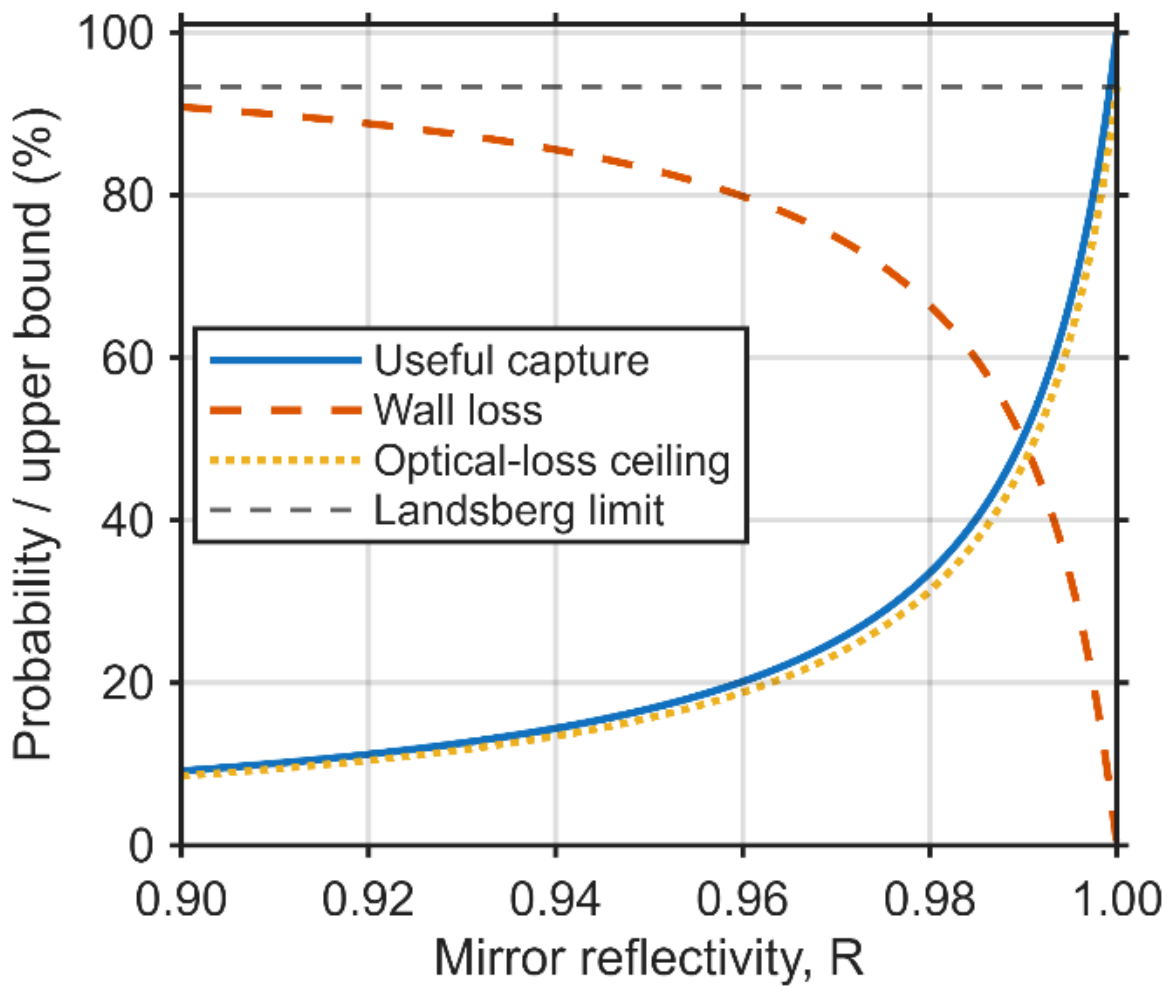

FIG 6. Photon-recycling enhancement of optical utilization for g=0.01. As the mirror reflectivity $R$ increases, the useful-conversion probability $P_{use}$ increases while the loss probability $P_{loss}$ decreases. The corresponding optical-loss-limited efficiency, $\eta_L P_{use}$, approaches the Landsberg limit $\eta_L = 93.3\%$ as $R \to 1$.